# Energy of Alfvén waves generated during magnetic reconnection

L. C. Wang[1], L. J. Li[1], Z. W. Ma[1]*, X. Zhang[1], L. C. Lee[2]

[1]Institute for Fusion Theory and Simulation, Zhejiang University, Hangzhou 310027, China.

[2]Institute of Earth Sciences, Academia Sinica, Taipei 11529, Taiwan

**Abstract:** A new method for the determination of the Alfvén wave energy generated during magnetic reconnection is introduced and used to analyze the results from two-dimensional MHD simulations. It is found that the regions with strong Alfvén wave perturbations almost coincide with that where both magnetic-field lines and flow-stream lines are bent, suggesting that this method is reliable for identifying Alfvén waves. The magnetic energy during magnetic reconnection is mainly transformed into the thermal energy. The conversion rate to Alfvén wave energy from the magnetic energy is strongly correlated to the magnetic reconnection rate. The maximum conversion rate at the time with the peak reconnection rate is found to be only about 4% for the cases with the plasma beta $\beta=0.01, 0.1$, and $1.0$.

*) Corresponding author: zwma@zju.edu.cn

## 1. Introduction

Space observations indicate that the temperature of the solar corona is of the order of multimillion degrees. It is much higher than the temperatures of the photosphere and the chromosphere, which are about 6000K and a few of tens of thousands degree, respectively. The coronal temperature rises rapidly in a narrow transition region between the chromosphere and the solar corona [1]. Since heat in the solar corona is lost through thermal conduction and radiation, there must be an effective energy supply and conversion mechanism that provides enough energy to the solar corona to maintain its high temperature. Solar coronal heating is still an outstanding problem in space physics.

At present, solar coronal heating mechanisms can be divided into two major categories: direct current heating and alternating current heating. The former can also be divided into two types: single-flux direct current models and multi-flux direct current models, both of which are closely related to magnetic reconnection [2]. There is also a synthetic heating mechanism [3] and shock heating mechanism [4]. For alternating current heating, or wave heating, the major problem is the transport and dissipation of energy, that is, how the energy is transported to the corona and dissipated. For the energy transport, large-scale Alfvén waves can play an important role [5]. Observations also showed that energy can be transported by Alfvén waves from the lower to the outer solar atmosphere and the energy flux is sufficient to heat the solar corona and accelerate the solar wind [6,7]. In recent years, many energy dissipation mechanisms have been proposed. In view of the high Lundquist number ($S \equiv \tau_R/\tau_A \sim 10^{10} - 10^{12}$, where $\tau_A$ and $\tau_R$ are the Alfvén and the magnetic diffusion times, respectively) and high magnetic fields in the solar corona, the wave energy can also be dissipated by the anisotropic viscosity instead of the resistivity [8]. Dissipation of the slow and fast MHD waves can also heat the corona [9]. In particular, fast magnetosonic waves are effective in perpendicular (to the magnetic field) proton heating [10]. Other works invoked Alfvén waves. For example, nonlinear Alfvén waves and Alfvén turbulence can play important roles in the energy dissipation [11-13]. It is therefore of interest to investigate in more detail the effect of

Alfvén waves on coronal heating.

Magnetic reconnection can be regarded as the most important process in the energy transfer and transport in astrophysical plasmas [14]. Through magnetic reconnection, magnetic energy can be transferred to the plasma thermal and kinetic energies. It is also widely recognized that magnetic reconnection is a fundamental energy transport mechanism in collisionless plasmas [15]. The plasma kinetic energy generated from the magnetic reconnection can be in the form of waves as well as bulk flow. Many waves, such as the Alfvén and magnetosonic waves, can be generated by magnetic reconnection [16-20]. Observed solar X-ray jets, which are thought to be a manifestation of the reconnection process, have been associated with the generation of Alfvén waves in the latter [21]. It is believed that magnetic reconnection takes place widely in the photosphere, the chromosphere and the solar corona [22]. Besides coronal heating, Alfvén waves are also important to solar wind acceleration. Hence, an investigation of the amount of Alfvén wave energy produced in magnetic reconnection is of significance.

In this Letter, we propose a method to determine the Alfvén wave energy in magnetic reconnection and carry out a resistive MHD simulation of the problem. In this method, the fast Fourier transform is used to determine the proportion of the Alfvén wave energy converted from the magnetic energy. We also examine the evolution of plasma kinetic energy, thermal energy, and the total released magnetic energy in the magnetic reconnection.

## 2. Simulation model

Our simulations are carried out in the two-dimensional domain $-L_x/2 \leq x \leq L_x/2, -L_z/2 \leq z \leq L_z/2$ and we assume $\partial/\partial y = 0$. Using $\mathbf{B} = \hat{\mathbf{y}} \times \nabla\psi + B_y\hat{\mathbf{y}}$, where $\psi$ is the flux function, the resistive MHD equations become [23]

$$\frac{\partial \rho}{\partial t} = -\nabla \cdot (\rho\mathbf{v}), \tag{1}$$

$$\frac{\partial (\rho\mathbf{v})}{\partial t} = -\nabla \cdot \left[\rho\mathbf{v}\mathbf{v} + \left(p + \frac{B^2}{2}\right)\overleftrightarrow{\mathbf{I}} - \mathbf{B}\mathbf{B}\right], \tag{2}$$

$$\frac{\partial \psi}{\partial t} = -\mathbf{v} \cdot \nabla \psi + \eta J_y, \tag{3}$$

$$\frac{\partial B_y}{\partial t} = -\nabla \cdot (B_y \mathbf{v}) + \mathbf{B} \cdot \nabla v_y + \eta \nabla^2 B_y, \tag{4}$$

$$\frac{\partial \varepsilon}{\partial t} = -\nabla \cdot \mathbf{Q}, \tag{5}$$

where energy density $\varepsilon$ and energy flux $\mathbf{Q}$ in (5) are given by

$$\varepsilon = \frac{\rho v^2}{2} + \frac{B^2}{2} + \frac{p}{\gamma - 1}, \tag{6}$$

$$\mathbf{Q} = \left(\frac{\rho v^2}{2} + B^2 + \frac{\gamma p}{\gamma - 1}\right)\mathbf{v} - (\mathbf{B} \cdot \mathbf{v})\mathbf{B} + \eta \mathbf{J} \times \mathbf{B}. \tag{7}$$

In (1)-(7), $\mathbf{v}, \mathbf{B}, \rho, p, \overleftrightarrow{\mathbf{I}}$ are the plasma velocity, the magnetic field, the plasma density, the thermal pressure, and the unit tensor, respectively. The variables $\mathbf{x}, \mathbf{v}, t, \mathbf{B}, \rho, p, \psi$ are normalized as follows: $\mathbf{B}/B_0 \to \mathbf{B}$, $\mathbf{x}/a \to \mathbf{x}$, $t/\tau_A \to t$, $\mathbf{v}/v_A \to \mathbf{v}$, $\psi/(B_0 a) \to \psi$, $\rho/\rho_0 \to \rho$, $p/(B_0^2/4\pi) \to p$, where $v_A$ is the Alfvén velocity, $\rho_0$ is the constant mass density, $a$ is the scale length, $\tau_A = a/v_A = a(4\pi\rho_0)^{1/2}/B_0$ is the Alfvén time, $\eta$ is the resistivity of the plasma, and $\gamma = 5/3$ is the ratio of the specific heats of the plasma.

The initial equilibrium magnetic field is given by

$$B_x = B_i \tanh\left(\frac{z}{\lambda_B}\right), B_y = 0, B_z = 0, \tag{8}$$

and the initial mass density profile is given by

$$\rho = \rho_i, \tag{9}$$

where $B_i$ and $\rho_i$ are positive constants. $\lambda_B$ is the half width of the current sheet in the initial geometry. The initial equilibrium is force-balanced. The thermal pressure is obtained by solving the equilibrium equations and is given by

$$p = (1 + \beta_\infty)\frac{B_i^2}{2} - \frac{B^2}{2}, \tag{10}$$

where $\beta_\infty$ is the plasma beta far away from the initial current sheet, where we have assumed that the initial state is static. In order to achieve fast reconnection, we shall use for the plasma resistivity

$$\eta = \eta_0 + \eta_l \exp[-(x^2 + z^2)], \tag{11}$$

where $\eta_0$ is the background resistivity and $\eta_l > \eta_0$ is the local resistivity.

A Runger-Kutta scheme with a fourth-order accuracy in time and space is employed to solve the resistive MHD equations. The size of simulation box is set to be $L_x = 32$ and $L_z = 16$. For the other variables, we choose $B_i = 1.0$, $\lambda_B = 1.0$, $\rho_i = 1.0$, and $\beta_\infty = 0.1$. We use 1280 grid points with uniform spacing along $x$ axis and 880 non-uniform grid points spacing along $z$ axis. The smallest grid spacing along $z$ axis is 0.002. We set the background resistivity to be $\eta_0 = 1 \times 10^{-4}$ and the local enhanced resistivity to be $\eta_l = 2 \times 10^{-3}$. In the $x$ direction, we use the periodic boundary condition. In order to guarantee the conservation of total energy, the closed boundary condition is used in the $z$ direction.

## 3. Perturbations of magnetic field and plasma flow during reconnection

Before examining Alfvén wave energy generated during magnetic reconnection, we first present perturbations of magnetic field and plasma flow at the time the reconnection rate reaches its maximum.

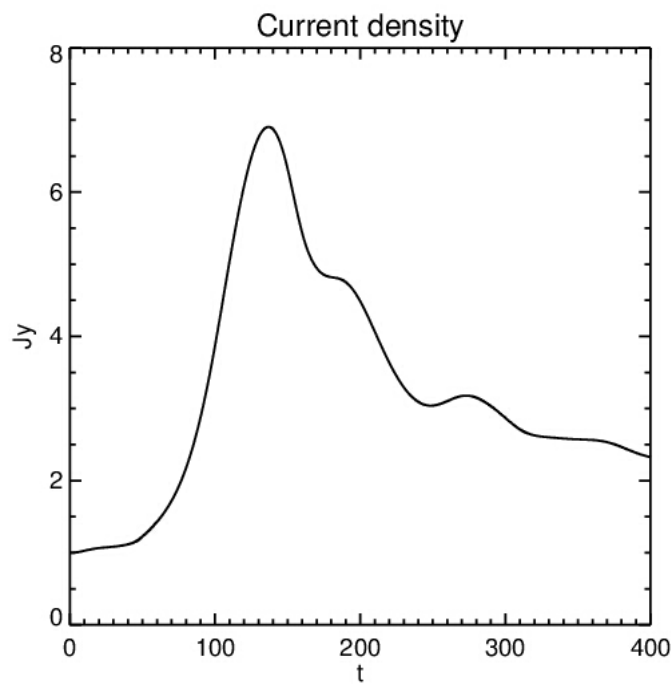

Figure 1. The current density at the X point as a function of time.

Figure 1 shows the current density at the X-point as a function of time. It can be seen that, in the early stage, reconnection rate exhibits an exponential growth. It then starts to fall slowly compared to the boost stage. Figure 2 shows the color contour plot of the perturbed $\delta B_x = B_x - B_{x0}$ and $U_x$ ($\mathbf{U} = \sqrt{\rho}\mathbf{v}$) at the time with the maximum reconnection rate. It is evident that the main perturbations of magnetic field and flow velocity take place in different regions. For example, on the *x* axis, the magnitude of $\delta B_x$ is zero while the magnitude of $U_x$ is at its maximum. As we know, the magnetic field and plasma flow perturbations associated with Alfvén wave should satisfy the Walen relation in the normalized form $\delta \mathbf{B} = \pm\sqrt{\rho}\delta\mathbf{v},$ which suggests that each component of the magnetic field and plasma flow perturbations in the same location must have the same magnitude and spatial scale with the in- or anti-phase.

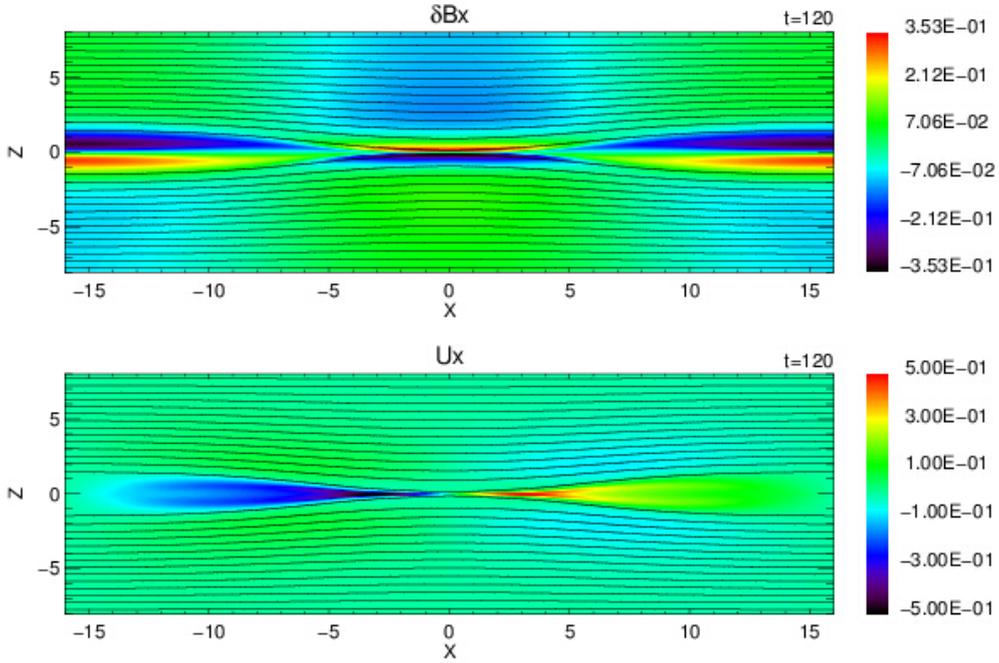

Figure 2. Contour plots of the perturbed $\delta B_x$ (top) and $U_x$ (bottom).

### 4. A method to determine the Alfvén wave energy

In this section, we develop a new method to identify the power of Alfvén waves through two steps: (1) A forward Fourier transformation of the perturbations of magnetic field $\delta \mathbf{B} = \mathbf{B} - \mathbf{B}_0$ and plasma flow $\mathbf{U}$ to the wave number **k** space and the spectral power selection, and (2) An inverse Fourier transformation of $\delta\widetilde{\mathbf{B}}$ and $\widetilde{\mathbf{U}}$ to the configuration space and the amplitude selection.

Step 1. Selection of spectral powers of magnetic field perturbation and plasma flow perturbation in the Fourier space. A forward Fourier transformation on

$\delta \mathbf{B} = \mathbf{B} - \mathbf{B}_0$ and $\mathbf{U}$, components by components, is carried out to get the spectral powers of magnetic field and flow perturbations in the wave number $\mathbf{k}$ space. From the Walen relation, it is known that the magnetic field and plasma flow perturbations should have the same magnitudes for all wavelengths, i.e., the spectral powers should be the same, if the perturbation is of an Alfvén wave. Thus, by choosing the smaller value of the spectral powers of magnetic field and plasma flow perturbations in the Fourier space, we can determine the magnitude of the magnetic field and plasma flow perturbations which may be associated with the Alfvén wave perturbations.

In the Fourier space, the phases associated with $\delta B_x$ and $U_x$ could be different. Figure 3 shows Type I and II distributions for the magnetic field and plasma fluid perturbations after the selection of the magnitudes of magnetic field and flow in Step 1. In Type I, the original phase of $\delta B_x$ is used, while the original phase of $U_x$ is adopted in Type II. It is evident that the two-type distributions do not match with each other as we expect. The mismatch is associated with the observation that in the Fourier space the phases of the perturbations of the magnetic field and the plasma flow are not the same. Thus, both Type I and Type II do not exactly represent the perturbations of the magnetic field and the plasma flow associated with Alfvén waves.

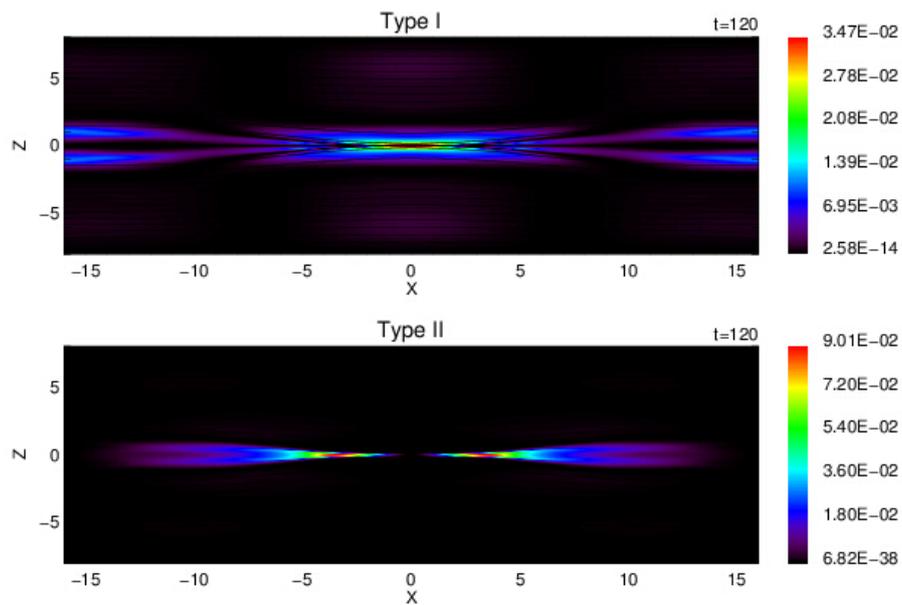

Figure 3. The perturbation distributions after spectral power selection. Type I and II distributions are based on the original $\delta B_x$ and $U_x$ phases, respectively

Step 2. Selection of magnitudes of magnetic field and flow perturbations after inverse Fourier transformation. Since we are not able to determine their phases in the Fourier space, but we know that the original phases or locations of the perturbations should not be changed, which suggests that its original phases should be used for the perturbations of the magnetic field and the plasma flow. By combining the selected spectral powers with the original phases, we are able to obtain the distributions of magnetic field and flow perturbations in the configuration space after the inverse Fourier transforms are performed as shown in Figure 3. We should note that for Alfvén waves, the distributions of magnetic field and flow perturbations in the configuration space have the same spectral power. Although these two spatial distributions look quite different, the perturbations of the magnetic field and the plasma flow overlap in some areas. Thus, it is natural to assume that the overlapped parts of the distributions are from Alfvén wave perturbations. Accordingly, we select the Alfvén wave perturbations by choosing the smaller value of the amplitudes (as given in Figure 3) at every point.

Figure 4 shows the finally selected distributions of Alfvén wave perturbations in the configuration space. It is shown that the $x$ component of the Alfvén wave is mainly located in the two sides of the outflow region. The distribution of the Alfvén wave energy looks like a crab's claw. But in the vicinity of the $x$ axis, although the flow velocity $v_x$ is the largest, there is no Alfvén wave due to the plasma velocity being perpendicular to magnetic field. Another region with strong Alfvén wave perturbation is around the separatrices. The bifurcated narrow strips of the Alfvén wave perturbation result from that the plasma flow suddenly changing direction, as shown in Figure 5. The gap between these two regions of the Alfvén wave perturbation is associated with the zero perturbation of the magnetic field $B_x$.

For the $z$ component of the Alfvén wave perturbation, it is found that they exist in both the inflow and outflow regions. In the outflow region, the distribution and location of this component of the Alfvén wave perturbation are quite similar to that of the $x$ component. But, it should be noted that the magnitude for the $z$ component of the Alfvén wave perturbation is much smaller than that for the $x$ component of the

Alfvén wave perturbation. The Alfvén wave perturbations in the inflow region are associated with the inward flow of magnetic reconnection and the flow from magnetic island expansion. Actually, this type of the Alfvén wave perturbations also exists for the $x$ component. Because the magnitude of the $x$ component in the inflow region is too small to compare with that in outflow region, it becomes invisible.

From Figure 5，we see that the regions with the strong Alfvén wave perturbations are coincident to the regions where both the magnetic field lines (yellow lines) and flow streaming lines (red lines) are bent. In some regions, the Alfvén wave perturbation is absent or weak, even if both magnetic field lines and flow streaming lines are strongly bent. This is because either the magnetic field or plasma flow is too weak, or the magnetic field and the plasma flow are perpendicular to each other. It is thus reasonable to conclude that the Alfvén wave selected by our method is reliable.

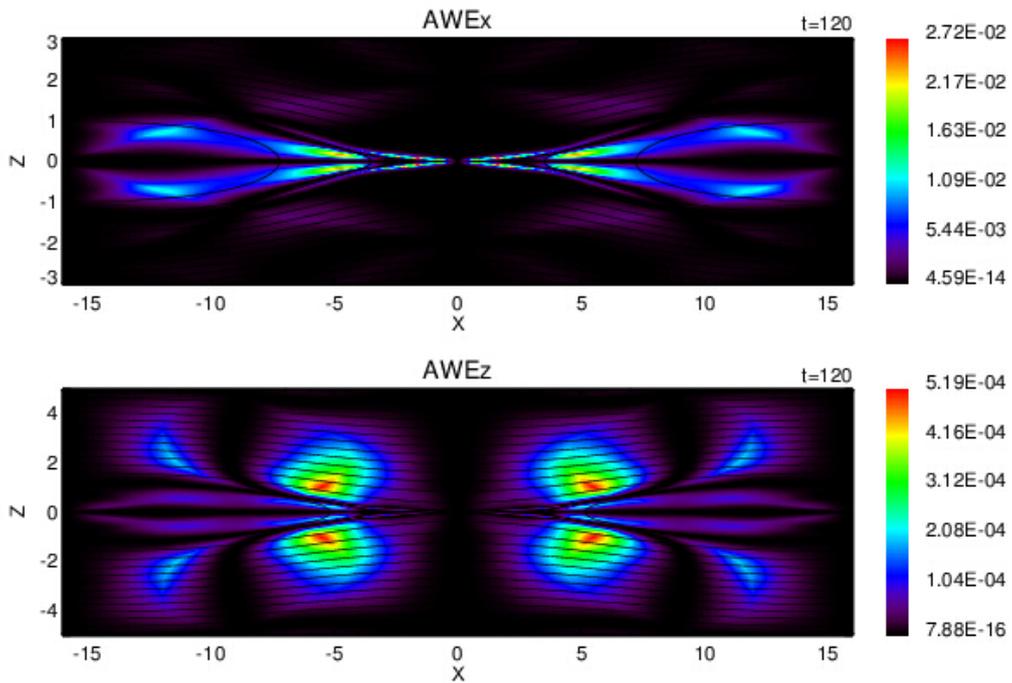

Figure 4. The energy distribution for the $x$ and $z$ components of Alfvén waves in the configuration space.

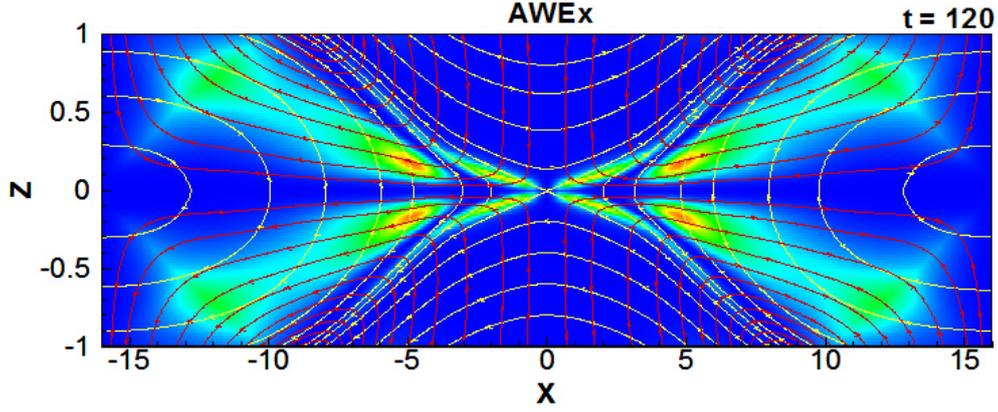

Figure 5. Detailed energy distributions for $x$ component of Alfvén waves with stream lines (yellow) and magnetic field lines (red).

## 5. The conversion rate to Alfvén wave energy from magnetic energy

We now use our selection method to investigate the generation of Alfvén waves in magnetic reconnection. Figure 6 shows the time variations of the magnetic energy (ME), thermal energy (TE), kinetic energy (KE) and Alfvén wave energy (AWE). $\text{AWE} = \iint \left[\frac{(\delta B)^2}{2\mu_0} + \frac{\rho v^2}{2}\right] dxdz = 2 * \text{KE}_{\text{awe}}$, where $\delta B$, $\sqrt{\rho}v$, and $\text{KE}_{\text{awe}}$ are the magnetic field perturbation, the plasma flow perturbation, and the kinetic energy associated with an Alfvén wave, respectively. It is obvious that ME decreases while TE, KE and AWE increase during magnetic reconnection. ME mainly transforms into TE and only a small part of ME goes into KE. According to the process of reconnection, ME always decreases and TE always increases. Figure 7 shows conversion rate from ME to TE, KE, and AWE. It is quite evident that the energy conversion rate between ME and KE is directly related to the rate of magnetic reconnection.

If we assume the whole KE generated during magnetic reconnection is associated with Alfvén wave perturbations, the maximum conversion rate from ME to AWE will be up to 2*KE~30%. But it is obvious that this large conversion rate is largely overestimated because the majority part of the jet flow in the outflow region is not related to Alfvén wave perturbations. After the selection of the spectral powers of perturbations in the Fourier space, we find that the magnetic field perturbation resulted from compression wave in the inflow region and the plasma jet flow in the outflow region are not eliminated yet. After both the spectral power selection in the Fourier space and the amplitude selection after the inverse Fourier transformation, it is found that the maximum conversion rate from ME to AWE largely drops to around 4% only. It should be pointed out that the 4% conversion rate could be underestimated because the selected magnetic field or plasma flow perturbations associated with an Alfvén wave could be reduced by the presence of a magnetosonic wave or a sonic wave or both waves. Thus, the 4% conversion rate from ME to AWE should be a lower bound. But we believe that the real figure of Alfvén wave energy will not be deviated much from the 4% conversion rate that has been cross-examined with the properties of the magnetic field lines and the flow stream lines in Figure 5. We also conduct several runs with the plasma beta from 0.01 to 1.0. It is found that the maximum conversion rate 4% from ME to AWE is almost independent of the plasma beta.

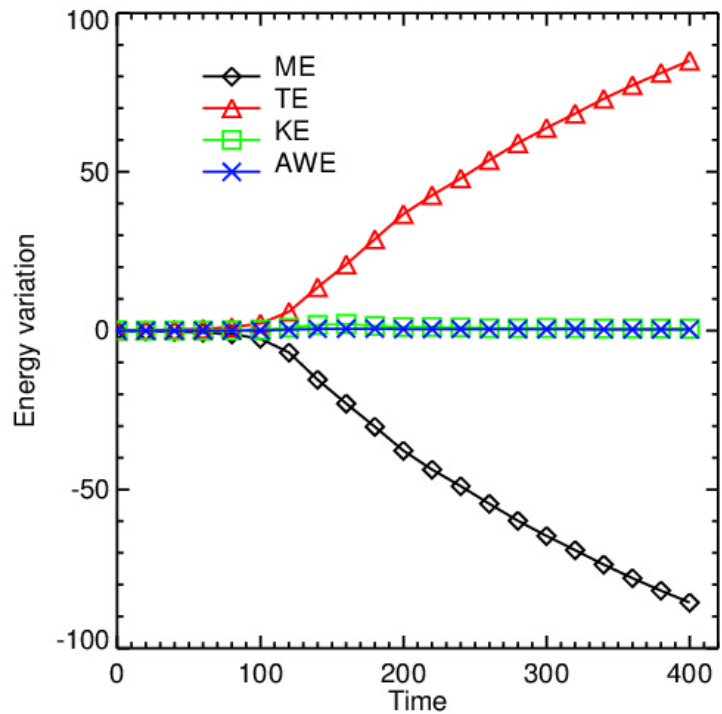

Figure 6. The time variations of the thermal energy (TE), the kinetic energy (KE), the Alfvén wave energy (AWE), and the magnetic energy (ME).

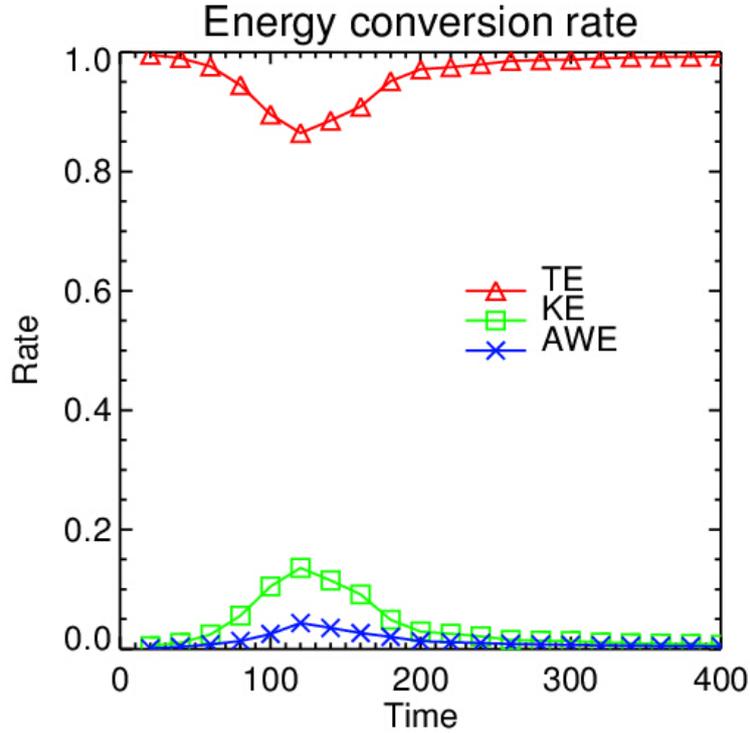

Figure 7. Conversion rate from magnetic energy to thermal energy (TE), to kinetic energy (KE), and Alfvén wave energy (AWE).

## 6. Conclusions

In this Letter, a new method for the determination of Alfvén wave energy during magnetic reconnection is introduced. It is used to analyze data obtained from two-dimensional MHD simulations of magnetic reconnection through the Fourier transformation and the spectral power selection as well as the inverse Fourier transformation and the amplitude selection. It is found that the regions with strong Alfvén wave perturbations are coincident with the regions where both magnetic field lines and flow stream lines are bent. Through the power and amplitude selections, magnetic field perturbations resulted from compression waves in the inflow region and plasma flow perturbation associated with the plasma jet flow in the outflow region can be eliminated from the Alfvén waves. It may be concluded that the Alfvén wave perturbation obtained by our method is a reasonable estimate. During magnetic reconnection, magnetic energy is mainly transformed into thermal energy, and the maximum Alfvén wave energy converted from magnetic energy is only about 4% for the cases with the plasma beta $\beta=0.01, 0.1$, and $1.0$.

**Acknowledgement**

This work is supported by Fundamental Research Fund for Chinese Central Universities, the National Natural Science Foundation of China under Grant No. 11175156 and 41474123, the ITER-CN under Grant Nos. 2013GB104004 and 2013GB111004, and by Ministry of Science and Technology in Taiwan under grant 101-2628-M-001-007-MY3.